\documentclass{article}
\usepackage{spconf,graphicx}
\usepackage{multirow}
\usepackage{tikz}
\usepackage{booktabs}
\usepackage{csquotes}
\usepackage{multirow}
\usepackage{caption,subcaption}
\usepackage{enumitem}

\title{AUDIO-VISUAL CHILD-ADULT SPEAKER CLASSIFICATION IN DYADIC INTERACTIONS}

\name{Anfeng Xu$^{\star}$,\; Kevin Huang$^{\star}$,\; Tiantian Feng$^{\star}$,\; Helen Tager-Flusberg$^{\dagger}$,\; Shrikanth Narayanan$^{\star}$\thanks{This work was supported by funds from Apple.}}

\address{$^{\star}$ University of Southern California, Los Angeles, CA, USA \\
  $^{\dagger}$ Boston University, Boston, MA, USA }

\begin{document}
\ninept
\maketitle
\begin{abstract}
\end{abstract}
%

Interactions involving children span a wide range of important domains from learning to clinical diagnostic and therapeutic contexts. Automated analyses of such interactions are motivated by the need to seek accurate insights and offer scale and robustness across diverse and wide-ranging conditions. Identifying the speech segments belonging to the child is a critical step in such modeling. 
Conventional child-adult speaker classification typically relies on audio modeling approaches, overlooking visual signals that convey speech articulation information, such as lip motion.
Building on the foundation of an audio-only child-adult speaker classification pipeline, we propose incorporating visual cues through active speaker detection and visual processing models. Our framework involves video pre-processing, utterance-level child-adult speaker detection, and late fusion of modality-specific predictions.
We demonstrate from extensive experiments that a visually aided classification pipeline enhances the accuracy and robustness of the classification.
We show relative improvements of $2.38\%$ and $3.97\%$ in F1 macro score when one face and two faces are visible, respectively.


\begin{keywords}
speaker classification, child speech, audio-visual, deep learning, autism
\end{keywords}

\section{Introduction}
\label{sec:intro}

Many important child-centric applications, such as neurocognitive disorder assessment, psychotherapy interventions, child forensic interviews in the legal domain, and personalized learning, involve direct, often dyadic, interactions between the child and an adult expert, e.g., clinician, caregiver, or teacher.  These interactions serve as a means for understanding children's behavior in the target contexts leading to specific screening, diagnostic, and treatment measures. 
The primary area of focus for this work is Autism Spectrum Disorder, a neurocognitive developmental disorder that is increasingly diagnosed among children, underscoring an imminent need for broad access to precise and early assessment. In the United States, 1 in 36 children are estimated to be diagnosed with Autism Spectrum Disorder in 2023 \cite{maenner2023prevalence}. Early diagnosis and interventions are crucial for positive, long-term developmental outcomes. The assessment of Autism Spectrum Disorder has primarily relied on dyadic interaction sessions such as ADOS \cite{lord2000autism} and ELSA \cite{barokova2021eliciting}. Analysis of children's spoken language is particularly beneficial for identifying children with Autism Spectrum Disorder \cite{sorensen2019cross}, who are typically delayed in acquiring language and communication skills. 

Computational methods enable data-driven approaches to support Autism Spectrum Disorder assessments through automated analysis of interactions with the child, substantially enhancing clinical and research outcomes. The child-adult speaker classification task is a critical component in these data-driven analyses. This task involves partitioning dyadic interaction recordings into child and adult speech segments, allowing efficient and accurate speech feature extraction that helps understanding interaction dynamics. For example, prior studies have shown that the amount of intelligible child speech, which can be derived following the child-adult speaker classification, indicates the child's language capabilities \cite{xu23e_interspeech}. Similarly, conversational latency from children is known to be a robust marker as a treatment response in Autism Spectrum Disorder \cite{mckernan2022intra}. 

Conventional child-adult speaker classification primarily adopts audio-only modeling approaches. However, the audio modality may frequently be insufficient for identifying speaker identities in child-adult interactions, especially when the quality of the audio signal is compromised. In addition, high variability in children's speech production makes the audio-based disambiguation of speakers even more challenging. Finally, data from these natural, spontaneous interactions are also characterized by extensive variability in the ambient conditions of the interaction.

A major source of additional information that could assist child-adult speaker classification is dynamic visual cues about the speaker's articulators (e.g., lips, jaw), providing direct measures for inferring who is speaking during the dyadic interactions. There have been advances supporting the visual speech processing of children, including the creation of a large, balanced child face corpus \cite{karkkainen2019fairface} and improved active speaker detection (ASD) models. The natural extension in child-adult speaker classification is to leverage cues from visual modality.


\begin{figure}[!t]

\begin{minipage}[b]{1\linewidth}
  \centering
  \vspace{-4mm}
  \centerline{\includegraphics[width=0.9\textwidth]{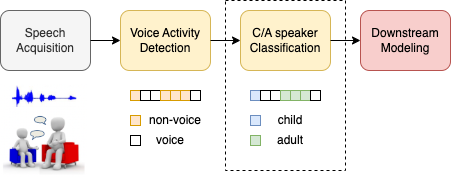}}
  \vspace{-4mm}
  \caption{Spoken language assessment pipeline}\medskip
  
\end{minipage}
\vspace{-9mm}
\end{figure}

\begin{figure*}
    \centering
    \includegraphics[width=0.95\linewidth]{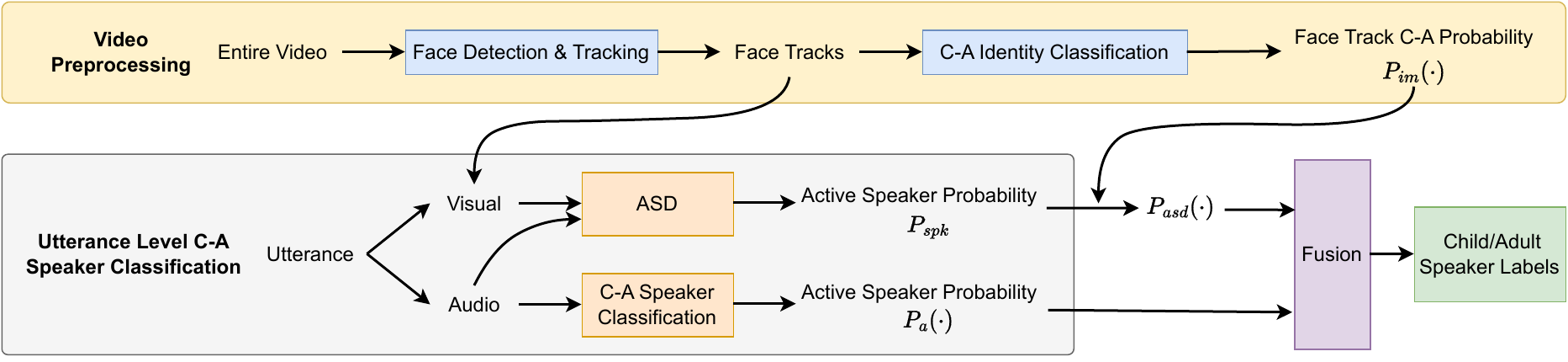}
    \vspace{-2mm}
    \caption{Proposed visual-assisted multimodal child-adult speaker classification in dyadic interactions.}
    \label{fig:audio_visual_pipeline}
    \vspace{-3mm}
\end{figure*}

In this paper, we propose a visually-assisted child-adult classification pipeline that builds on the foundation of an audio-only approach.
We use the WavLM \cite{chen2022wavlm} and Whisper \cite{radford2023robust} encoders for the audio-only modeling, ViT \cite{dosovitskiy2020image} and ResNet \cite{he2016deep} for the visual feature processing, and Light-ASD \cite{liao2023light} for active speaker detection.
Our main contributions are summarized below.

\begin{itemize}[leftmargin=*]
  \item  We introduce one of the first visually-aided child-adult speaker classification pipelines. Specifically, the proposed framework integrates recent advances in active speaker detection (ASD) modeling to assist the existing audio-only speaker classifier.
  
  \item To validate our proposed modeling framework, we conduct a novel annotation process that provides labels with speaker identities based on child-adult interaction videos, offering approximately 30,000 utterances for the multimodal modeling experiment.
  \item Our experiments on clinical data from child-caregiver interactions in the context of autism demonstrate that the proposed multimodal child-adult speaker classification pipeline can provide substantial improvements in identifying the speaker of interest from child-adult dyadic interactions. Specifically, we show relative improvements of approximately $2.4\%$ (1 visible face) and $4.0\%$ (2 visible faces) in F1-macro scores.
\end{itemize}

\section{BACKGROUND}
\subsection{Child-adult speaker classification}

Earlier work includes using i-vectors and x-vectors for diarization involving children \cite{najafian2016speaker, cristia2018talker}. However, conventional speaker diarization approaches using clustering methods are not suitable to our context since they rely on a substantial amount of linguistically well-formed speech segments from every speaker, which are often limited in child-centric clinical domains. Child-adult classification is an alternative framework to segment child and adult speech regions.  Koluguri et al.~\cite{koluguri2020meta} investigated the use of meta-learning on x-vectors for the child-adult speaker classification task. More recently, Self-Supervised Learning (SSL) models and the Whisper encoder have been shown to yield remarkably reliable performance on the child-adult speaker classification task \cite{xu23e_interspeech, lahiri23_interspeech}.

\subsection{Active speaker detection}

In the visual domain, active speaker detection (ASD) refers to the task of identifying the speaker's face in a visual scene. The large-scale dataset AVA-Active Speaker \cite{roth2020ava} introduced in 2019 has advanced the development of numerous ASD models. 
In this work, we use Light-ASD~\cite{liao2023light}, a recently introduced model 
that achieves competitive performance on the AVA-Active Speaker benchmark while requiring a relatively simple architecture with fewer model parameters, making it an ideal modeling framework for accommodating our target application. While the original ASD framework aims at predicting the speaker at the frame level, we adapt Light-ASD to predict the speaker per utterance.

\section{METHOD}

Fig~\ref{fig:audio_visual_pipeline} shows our proposed audio-visual child-adult classification framework, which consists of three parts: video pre-possessing, utterance-level child-adult speaker detection, and late fusion.
Given a video $\mathbf{V}$ and the starting and ending timestamps from the $k^{th}$ utterance in the video $t_s^k$ and $t_e^k$, our aim is to map $v_{t_s^k:t_e^k} \rightarrow y^k$. Here, $v_{t_k^s:t_k^e}$ corresponds to the video segment from time $t_s^k$ to $t_e^k$ and $y^k \in \{0, 1\}$ corresponds to the speaker identity (adult or child).

\subsection{Video pre-processing}
We use the S$^3$FD \cite{zhang2017s3fd} method for face detection and tracking. The $N$ detected face tracks $Tr^1, Tr^2, ..., Tr^N$ with the corresponding timestamps $(\tau_s^1, \tau_e^1), ..., (\tau_s^N, \tau_e^N)$ are passed into a child-adult image classifier. We use ViT-Base and ResNet-34 for the image classification task and obtain the child probability $P_{im}(c)$ for each face track. We only consider the utterances for which either or both of the child and adult face tracks satisfy $\tau_s^i \leq t_s^j$ and $t_e^i \leq \tau_e^j$, where $i$ and $j$ are indices for the utterances and face tracks, respectively. For simplicity, we will omit the indices in the rest of the paper.

\subsection{Utterance-level speaker classification}
\subsubsection{Light-ASD adaptation}
\label{subsec:asd_adapt}

Light-ASD~\cite{liao2023light} uses convolutional layers to extract visual and audio features. To make the model lightweight, it uses 2D convolution (spatial) followed by 1D convolution (temporal) for the visual input and two 1D convolutions (MFCCs and temporal) for the audio input. For both visual and audio encoders, Light-ASD combines two separate convolutional layer paths with different kernel sizes.

\begin{figure}
    \centering
    \includegraphics[width=0.75\linewidth]{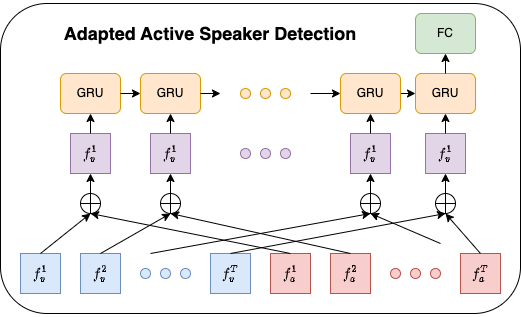}
    \vspace{-2.5mm}
    \caption{Proposed Light-ASD adaptation outputs utterance-level speaker probabilities.}
    \label{fig:lasd}
    \vspace{-3mm}
\end{figure}


\begin{figure}[htb]
    \centering
    \includegraphics[width=0.7\linewidth]{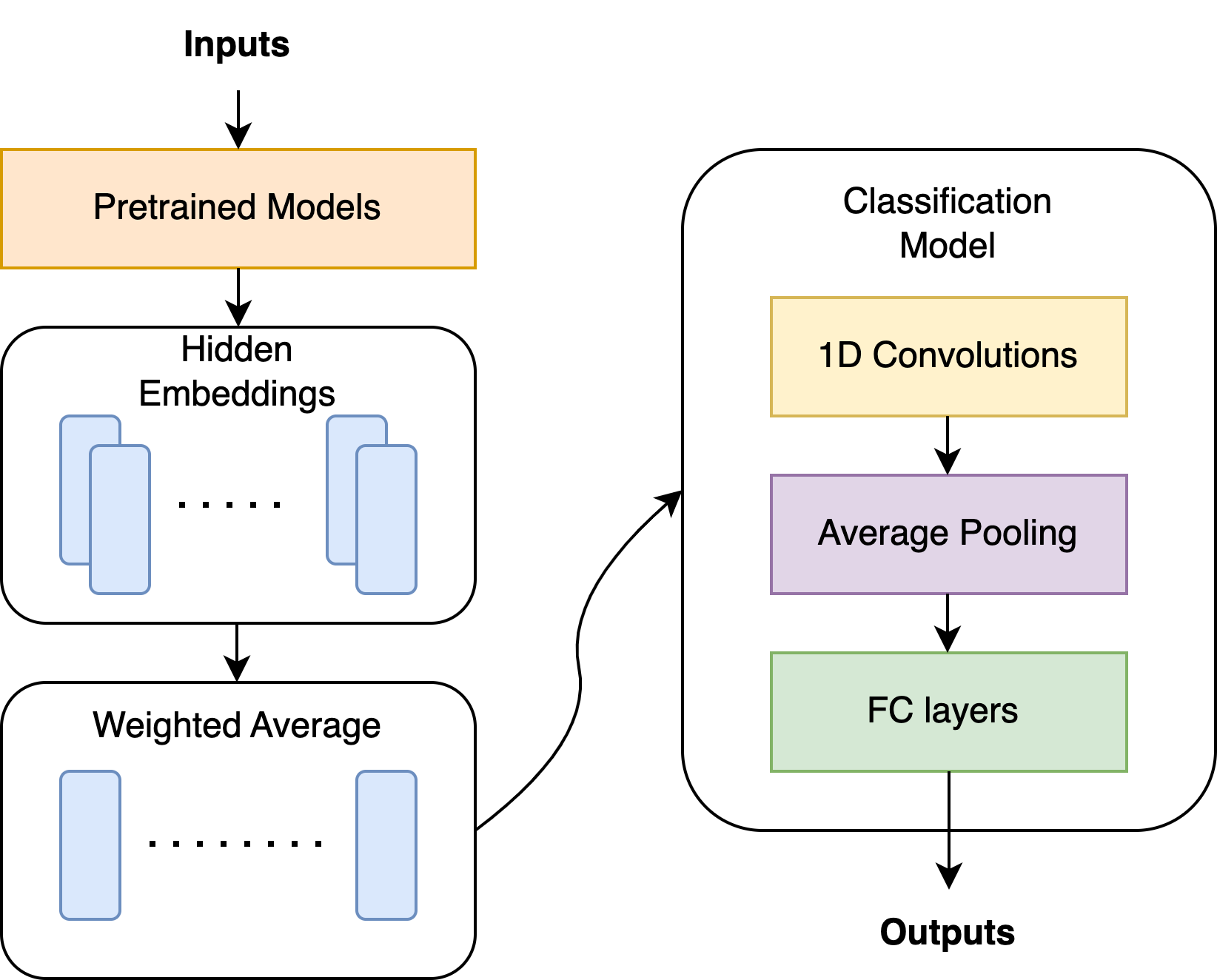}
    \vspace{-2.5mm}
    \caption{Audio-only child-adult speaker classification model.}
    \label{fig:audio}
    \vspace{-3mm}
\end{figure}

Light-ASD uses a bidirectional GRU to predict if the candidate speaks in each frame, but we use an unidirectional GRU to predict if the candidate speaks per utterance as in Fig~\ref{fig:lasd}. While the original Light-ASD uses a visual auxiliary classifier that builds FC layers directly on top of the visual embeddings, we similarly insert an FC layer after unidirectional GRUs to predict utterance-level speaker labels. Similar to the original Light-ASD, we use the binary cross entropy loss where the combined loss is $L = L_{av} + 0.5L_v$. Here, $L_{av}$ and $L_v$ are the losses from the audio-visual classifier and visual auxiliary classifier, respectively. At inference time, only the audio-visual classifier is active. If only either the child or adult face track is available, the adapted Light-ASD outputs the speaking probability $P_{spk}$. When both child and adult faces tracks are available, we propose to compute the two speaker probabilities $P_{spk}^1$ and $P_{spk}^2$ from ASD using the following two approaches:

\vspace{1mm}
\noindent \textbf{Individual input:} We can pass the two visual embeddings $f_{v_1}^1 ... f_{v_1}^T$, $f_{v_2}^1 ... f_{v_2}^T$ with the same audio embeddings $f_a^1 ... f_a^T$ separately to the adapted Light-ASD to get $P_{spk}^1$ and $P_{spk}^2$ and normalize such that $P_{spk}^1 + P_{spk}^2 = 1.$ We want to highlight that the adapted Light-ASD uses individual input in one-face condition.

\noindent \textbf{Combined input:} We propose to concatenate the two inputs, where the visual input is $[f_{v_1}^1, f_{v_2}^1] ... [f_{v_1}^T, f_{v_2}^T]$ and audio input is $[f_a^1, f_a^1] ... [f_a^T, f_a^T]$, so that the model can predict $P_{spk}^1$ and $P_{spk}^2$ directly. This way, the model can learn to compare the two visual inputs instead of making predictions independently.

\vspace{-0.2cm}
\subsubsection{Audio-only child-adult speaker classification}
Each utterance is passed to a pre-trained speech embedding model and the hidden embeddings are extracted.
The hidden embeddings are averaged with learnable weights.
Then, the weighted averaged embeddings are passed through a stack of three 1D convolution layers of hidden size 256 with a Rectified Linear Unit (ReLU) activation function.
The outputs are averaged over the timestamps, resulting in a single 256-dimensional embedding.
Finally, the child-adult probability is estimated by fully connected layers with one hidden layer of dimension 256. The pipeline is summarized in Fig~\ref{fig:audio}.

\subsection{Late fusion}
\label{sec:fusion}
\textbf{One-face ASD output:} As we described in Sec.~\ref{subsec:asd_adapt}, the adapted Light-ASD outputs the speaking probability $P_{spk}$ from the corresponding face track subset $Tr_{t_s:t_e}$ and $a_{t_s:t_e}$, where $a_{t_s:t_e}$ is the audio part of $v_{t_s:t_e}$. The visually guided child uttering probability can be calculated as follows: $$P_{asd}(c) = P_{spk} \cdot P_{im}(c) + (1 - P_{spk}) \cdot (1 - P_{im}(c))$$

\noindent \textbf{Two-faces ASD output:} When both child and adult face tracks are available, the visually-aided child probability is calculated as $$P_{asd}(c) = P_{im}^1(c) \cdot P_{spk}^1 + P_{im}^2(c) \cdot P_{spk}^2,$$ where $1$ and $2$ denote the first and second faces. The adult probability $P_{asd}(a)$ can be calculated similarly. 

\noindent \textbf{Audio and ASD fusion:} With the output from audio-only child-adult classification model $P_a(c)$ and $P_a(a)$, we determine the final prediction as follows:
\vspace{-1mm}
$$\hat{y} = argmax([P_{asd}(c) \cdot P_a(c), P_{asd}(a) \cdot P_a(a)])$$

\vspace{-2mm}
\section{DATASET}

\begin{table}[t]
  \centering
  \caption{Dataset statistics}
  \vspace{-3mm}
  \footnotesize
  \begin{tabular}{lc}
    \toprule
    \multicolumn{1}{c}{\textbf{Category}} & \textbf{Statistics}  \\
    \midrule
    Age (month) & Range: $49-95$ , Mean: $75.0$ , Std: $13.0$ \\
    Gender &  $68$ males, $19$ females \\
    Count per LL & $39$ (LL-1), $23$ (LL-2), $29$ (LL-3) \\
    Child Utterances Count & Range: $0-305$, Mean: $115.6$, Std: $74.0$ \\
    \bottomrule
  \end{tabular}
\vspace{-2mm}
\label{tab:demographics}
\end{table}

\begin{table}[t]
  \centering
  \caption{Utterance counts per visual condition}
  \vspace{-3mm}
  \footnotesize
  \begin{tabular}{lcccc}
    \toprule
    Visual condition & 0-Face & 1-Face (c/a) & 2-Faces (c\&a) & others\\

    Count & $5171$ & $9025$ & $13563$ & $443$ \\
    \bottomrule
  \end{tabular}
\vspace{-2mm}
\label{tab:utterances}
\end{table}

We use 87 Zoom recordings of dyadic activities between children and their parents from the dataset described in \cite{Butler, butler2022fine}. Experts have rated the language capabilities of children in these videos into three developmental levels: pre-verbal communication (LL-1), first words (LL-2), and word combinations (LL-3) from the transcripts. The audio was separately annotated into utterances with 4 labels, intelligible speech, unintelligible speech, nonverbal vocalization, and singing, in addition to the child-adult labels, as detailed in \cite{xu23e_interspeech}. The participants' information is reported in Table~\ref{tab:demographics}. 

Since very short utterances may introduce artifacts that challenge the identification of active speakers, we only evaluate utterances that are longer than 0.3 seconds. In addition, we truncate utterances longer than 3 seconds to 3 seconds, as most utterances have a shorter duration. This exclusion criterion leads to $10057$ child utterances and $18145$ adult utterances. We have also annotated each face track into child, adult, third person, and unclear. The numbers of available utterances for each visual scenario are shown in Table~\ref{tab:utterances}. The cases where a third person's face is included in the video or a detected face cannot be determined to be a child or adult are labeled as \enquote{others}. The videos are saved with $25$ fps, and the audio is recorded with a $16$ kHz sampling rate.

\section{EXPERIMENTAL SETUP}
\subsection{Video pre-processing}
To classify individual faces as depicting a child or an adult, we
use the pre-trained ViT-Base architecture with 16x16 patches from HuggingFace \cite{wolf2019huggingface} and the pre-trained ResNet-34 model from Torchvision \cite{marcel2010torchvision}. 
Both architectures are fine-tuned on the FairFace dataset \cite{karkkainen2019fairface} to predict the probability of a face lying in one of nine age brackets (e.g. $0$-$2$, $3$-$9$, etc). 
During evaluation on our dataset, we consider the first three age brackets as the child label and the latter six as the adult one. 
Faces from the FairFace dataset are cropped to be consistent with the cropping of Light-ASD, and then random augmentation is applied. 
Augmentation includes subsets of the following transformations: horizontal flip, changing brightness, contrast, hue, saturation, value parameters, blur, and random rectangular black occlusion. 
The ViT model is trained using the AdamW optimizer with learning rate $1\mathrm{e}{-6}$ and batch size 32 over 20 epochs. 
The ResNet model is trained using the Adam optimizer with learning rate $2\mathrm{e}{-7}$ and batch size 32 over 50 epochs. 
We average outputs from all face images to estimate the child-adult probability per track.

\subsection{Light-ASD Adapted Child-Adult Speaker Classification}

We perform 5-fold cross-validation with $60\%$ train, $20\%$ validation, and $20\%$ test splitting at the session level. We use the Adam optimizer with a learning rate of $1\mathrm{e}{-4}$, batch size of 64, and 20 epochs. 
For the individual visual input modeling, we use all utterances with at least the child or adult corresponding face. Only utterances with both child and adult face are used for the combined input modeling.

\subsection{Audio-only child-adult speaker classification}

The pre-trained embeddings are extracted using Huggingface. We perform 5-fold cross-validation with the same train-validation-test split as in the Light-ASD adaptation. We use Adam optimizer with a learning rate of $5\mathrm{e}{-5}$, batch size of 64, and 20 maximum epochs. 


\section{RESULTS AND ANALYSIS}

\begin{table}[!t]
  \centering
  \caption{Overall C-A classification (all visual conditions except others), F1 macro. We use combined input to the 2-Face condition.}
  \vspace{-3mm}
  \footnotesize
  \begin{tabular}{l c c c c}
    \toprule
    \multicolumn{1}{c}{\textbf{Modeling}}   & \textbf{Whisper-Base} &\textbf{Whisper-Small}  & \textbf{WavLM+} \\
    \cmidrule(lr){1-1} \cmidrule(lr){2-4}
    Audio & $84.2$ & $85.0$ & $84.9$ \\
    \cmidrule(lr){1-1} \cmidrule(lr){2-4}
    ASD + ResNet & \multicolumn{3}{c}{$76.4$ (Audio Model: N.A.)} \\
    ASD + ViT & \multicolumn{3}{c}{$76.9$ (Audio Model: N.A.)} \\
    \cmidrule(lr){1-1} \cmidrule(lr){2-4}
    Audio+ASD+ResNet & $\mathbf{86.7}$ & $\mathbf{87.4}$ & $\textbf{87.1}$ \\
    Audio+ASD+ViT & ${86.7}$ & ${87.3}$ & ${87.1}$ \\
    \bottomrule
  \end{tabular}
  \vspace{-2mm}
  \label{tab:results}
\end{table}

\begin{table}[!t]
  \centering
  \caption{Results by visual condition, F1 macro}
  \vspace{-3mm}
  \footnotesize
  \begin{tabular}{l c c c c}
    \toprule
    \multirow{2}{*}{\textbf{Modeling}} & \multirow{2}{*}{\textbf{0-Face}} & \multirow{2}{*}{\textbf{1-Face}} & \multicolumn{2}{c}{\textbf{2-Faces}} \\

     &  &  & \multicolumn{1}{c}{\textbf{Individual}} & \multicolumn{1}{c}{\textbf{Combined}} \\
    \cmidrule(lr){1-1} \cmidrule(lr){2-2} \cmidrule(lr){3-3} \cmidrule(lr){4-5}
    Audio & $\mathbf{84.0}$ & $84.1$ & $85.7$ & $-$ \\
    ASD + ResNet & $-$ & $70.9$ & $76.8$ & $79.9$ \\
    Audio+ASD+ResNet & $-$ & $\mathbf{86.1}$ & $\mathbf{88.9}$ & $\mathbf{89.1}$ \\
    \bottomrule
  \vspace{-5mm}
  \end{tabular}
  
  \label{tab:visual}
\end{table}

\begin{figure}[!t]
\begin{minipage}[b]{1\linewidth}
  \centering
  \centerline{\includegraphics[width=0.9\textwidth]{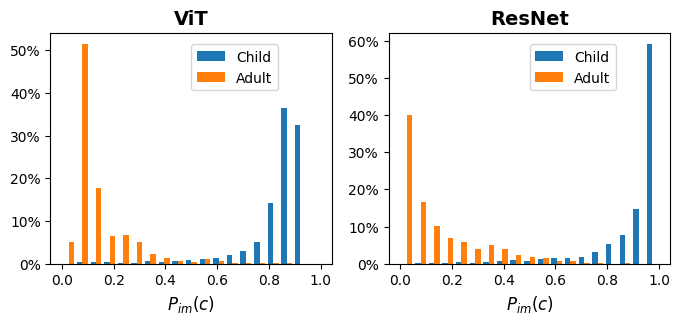}}
  \vspace{-4mm}
  \caption{Histograms of child probabilities from ViT and ResNet}\medskip
  \label{fig:histogram}
\end{minipage}
\vspace{-9mm}

\end{figure}

\subsection{Is the visually-aided approach beneficial?}
ViT and Resnet respectively result in $96.6\%$ and $96.2\%$ F1 macro scores in child-adult image classification, where the same label is assigned to all images within a face track. 
ResNet results in more confident probabilities for correctly classified samples as shown in Fig~\ref{fig:histogram}, resulting in a competitive multimodal performance.
The adapted Light-ASD results in $76.3\%$ F1 macro score with one input face for determining if the speaker is speaking and $83.4\%$ F1 macro score with two input faces for determining which face the speaker is.
The overall pipeline performances are summarized in Table~\ref{tab:results}.
We see improvements over the audio-only modeling approach in any combination of models we explored, while different audio and image models do not result in substantial differences in performance. The remaining paper compares speaker classification performances using Whisper-Small and ResNet as audio and image backbones.

\begin{table}[!t]
  
  \centering
  \caption{Results by utterance length (in seconds), F1 macro}
  \vspace{-3mm}
  \footnotesize
  \begin{tabular}{l c c c c c}
    \toprule

    \multirow{2}{*}{\textbf{Modeling}} & \multicolumn{1}{c}{\textbf{Visual}} & \multirow{2}{*}{\textbf{0.3-0.6}} & \multirow{2}{*}{\textbf{0.6-1}} & \multirow{2}{*}{\textbf{1-2}} & \multirow{2}{*}{\textbf{2-3}} \\

     & \multicolumn{1}{c}{\textbf{Condition}} & & & & \\
     
    \cmidrule(lr){1-1} \cmidrule(lr){2-6}

    Audio & & $75.6$ & $82.5$ & $88.4$ & $92.0$\\
    ASD + ResNet & 1-Face& $66.1$ & $66.5$& $73.6$ & $80.3$\\
    Audio+ASD+ResNet & & $\mathbf{78.0}$ & $\mathbf{85.1}$ & $\mathbf{89.9}$ & $\mathbf{93.6}$\\
    \cmidrule(lr){1-1} \cmidrule(lr){2-6}
    
    Audio & & $78.4$ & $84.3$ & $90.2$ & $94.2$\\
    ASD + ResNet & 2-Face & $75.0$ & $78.4$ & $82.4$ & $86.5$\\
    Audio+ASD+ResNet & & $\mathbf{83.4}$ & $\mathbf{88.6}$ & $\mathbf{92.5}$ & $\mathbf{94.8}$\\
    
    \bottomrule
  \end{tabular}
  \vspace{-2mm}
  \label{tab:length}
  
\end{table}

\begin{table}[!t]
  
  \centering
  \caption{Results by utterance types, F1 macro}
  \vspace{-3mm}
  \footnotesize
  \begin{tabular}{l c c c}
    \toprule
    \multirow{2}{*}{\textbf{Modeling}} & 
    \textbf{Visual} & 
    \multirow{2}{*}{\textbf{Speech}} & 
    \textbf{Nonverbal}  \\ 
     & \textbf{Condition} & & \textbf{Vocalization} \\
    \cmidrule(lr){1-1} \cmidrule(lr){2-4}
    Audio & & $84.4$ & $72.4$ \\
    ASD + ResNet& 1-Face & $72.2$ & $56.6$ \\
    Audio+ASD+ResNet & & $\mathbf{87.5}$ & $\mathbf{73.0}$ \\
    
    \cmidrule(lr){1-1} \cmidrule(lr){2-4}
    Audio & & $87.5$ & $71.7$ \\
    ASD + ResNet & 2-Faces & $81.7$ & $64.7$ \\
    Audio+ASD+ResNet & & $\mathbf{91.5}$ & $\mathbf{74.5}$ \\
    \bottomrule
  \vspace{-6mm}
  \end{tabular}
  
  \label{tab:type}
\end{table}

\subsection{How does the visual condition affect the results?}
We identify improvements of $2.38\%$ and $3.97\%$ in the F1 macro scores with the multimodal approach in 1-face and 2-face visual conditions, as demonstrated in Table~\ref{tab:visual}. This observation shows that the 2-face condition, where both child and adult faces are visible, provides more benefits to speaker classification by visual cues (e.g. articulation movements) from both sides for disambiguation of speech activities.
The combined input is used for the following analyses with the 2-face condition, as it shows superior performance.
\subsection{Does utterance duration impact speaker classification?}
We see that utterances with longer duration result in better child-adult speaker classification performances for both audio-only and visual-aid modeling frameworks, suggesting that longer segments provide more audio and visual cues that are useful to disambiguate speakers in dyadic interactions.
Encouragingly, the multimodal approach improves the performance more for shorter utterances as shown in Table~\ref{tab:length}, demonstrating the increased robustness of our proposed method with respect to utterance length.

\subsection{How challenging are non-verbal vocalizations in speaker classification?}
Overall, the classification of nonverbal samples is persistently lower than uttered speech samples across all experimental conditions.
We see the most classification improvements in uttered speech samples with visual aids, as shown in Table~\ref{tab:type}, while the improvements for nonverbal vocalization are less substantial.
This is likely because nonverbal vocalizations involve less specified visible articulation dynamics compared to uttered speech conditions, making Active Speaker Detection more challenging. 

\section{CONCLUSION}
\label{sec:conclusion}
In this work, we propose a visually aided multimodal framework for the child-adult speaker classification task for dyadic interactions. To this goal, we have annotated our dataset for speaker identity labels. Our experiments verify the hypothesis that incorporating visual cues improves 
the speaker classification task. Future work may include experimenting on additional child-adult dyadic interaction datasets, extending to non-dyadic scenarios, and integrating voice activity detection in the pipeline.

\bibliographystyle{IEEEbib}
\bibliography{strings,refs}

\end{document}